\documentclass[aps,prl,twocolumn,groupedaddress,showpacs]{revtex4}

\usepackage{amssymb,amsmath,graphicx}

\begin{document}

\title{Fine Structure of the Rydberg Blockade Zone}

\author{Yurii V. Dumin}
\email[]{dumin@yahoo.com,dumin@sai.msu.ru}
\affiliation{
Moscow State University,
GAISh, Universitetski pr.\ 13, 119992, Moscow, Russia}
\affiliation{
Space Research Institute (IKI) of Russian Academy of Sciences,
Profsoyuznaya str.\ 84/32, 117997, Moscow, Russia}

\date{December 30, 2013}

\begin{abstract}
A spatial structure of the zone blocked by the dipolar electric
field of a Rydberg atom is calculated taking into account
a possibility of excitation to the states with neighboring values
of the principal quantum number. As a result, it was found that
the blocked zone represents a number of co-centric spherical shells
rather than a solid ball, and the respective pair correlation
function should have additional maxima at small interparticle
distances.
\end{abstract}

%

\maketitle

The concept of Rydberg blockade of the ultracold gas began to be
widely discussed in atomic physics since the early 2000's~\cite{luk01},
and a few years later a possibility of this effect was verified
experimentally~\cite{ton04,sin04,wei09}.
In the resent time, the idea of Rydberg blockade is used in a lot of
intriguing phenomena, such as a highly efficient entanglement
between the light and atoms~\cite{li13,wei13}, observation of
the spatially ordered structures in a Rydberg gas~\cite{sch12},
realization of strong interaction between photons~\cite{pet12,wal12}
and, particularly, creation of the photon pairs~\cite{fir13,bos13},
\textit{etc.}

\begin{figure}
\includegraphics[width=8.5cm]{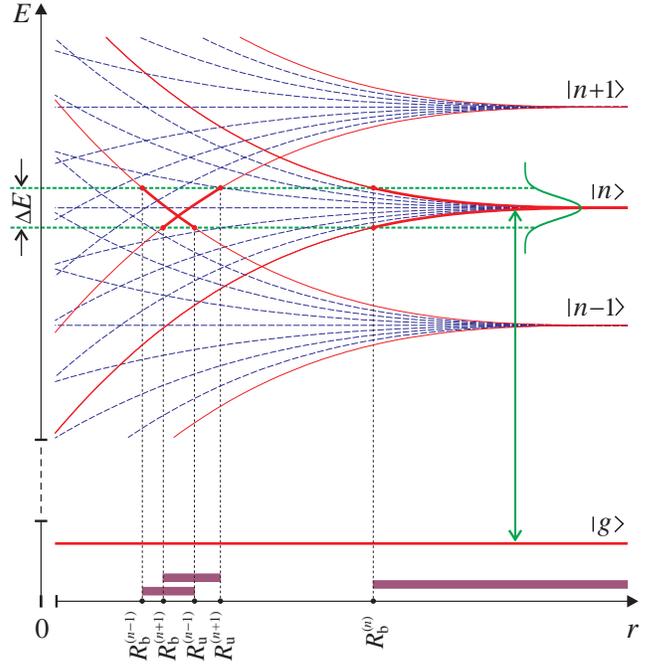}
\caption{\label{fig:Energy_levels}
Splitting the energy levels of a neighboring atom by
the electric field of the central Rydberg-excited atom.
Sublevels with maximal splitting are shown by solid (red)
curves; and other sublevels, by broken (blue) curves.
Rydberg excitation is allowed in the thick segments of
energy curves, located within the dotted (green) lines,
designating a characteristic bandwidth of the exciting
radiation. The corresponding intervals of radius
$ [R_{\rm b}^{(n-1)}, R_{\rm u}^{(n-1)}] $,
$ [R_{\rm b}^{(n+1)}, R_{\rm u}^{(n+1)}] $, and
$ [R_{\rm b}^{(n)}, +\infty ] $, where Rydberg excitation
can take place, are marked by thick (violet) strips near
the horizontal axis.}
\end{figure}

The basic mechanism of the Rydberg blockade can be illustrated in
Fig.~\ref{fig:Energy_levels}.
It is assumed here that Rydberg atom with principal quantum
number~$n$ located in the origin of coordinates ($ r=0 $)
produces an electric field which disturbs the energy levels of
neighboring atoms.
For example, in the simplest case of linear (with respect to
the field) Stark effect a highly-excited electron state
$ | n \rangle $ will be split into a symmetric series of sublevels,
characterized by the so-called parabolic quantum
numbers~\cite{gal94,yav80}. The degree of splitting evidently
increases as we approach the origin of coordinates, which is
shown in figure by a series of diverging curves.
Then, when a disturbed sublevel goes away from the characteristic
bandwidth of the exciting radiation~$ \Delta E $ (shown by a pair of
dotted horizontal lines), the excitation will no longer be possible.
Therefore, the ``excitation zone'' of the specified sublevel
should take place at $ r \geqslant R_{\rm b}^{(n)} $
(which is marked by the thick strip near the horizontal axis),
while at $ r < R_{\rm b}^{(n)} $ the Rydberg blockade develops~%
\footnote{
In fact, as is seen in Fig.~\ref{fig:Energy_levels},
the excitation of different sublevels will be blocked at
various radii~$r$. However, for the sake of definiteness,
we shall characterize the overall Rydberg blockade by
the behaviour of the most disturbed sublevel.}.

Unfortunately, the outlined standard picture of Rydberg blockade
does not take into account a presence of other energy levels with
close values of the principal quantum number, \textit{e.g.},
$ | n\!-\!1 \rangle $ and $ | n\!+\!1 \rangle $.
In fact, these energy levels experience a similar splitting
by the external electric field, so that the respective sublevels
can enter the energy band of the exciting radiation~$ \Delta E $
at the sufficiently small distances~$r$.
Namely, as illustrated in Fig.~\ref{fig:Energy_levels},
the most disturbed sublevel from the series $ | n\!-\!1 \rangle $
enters the band~$ \Delta E $ at the distance~$ R_{\rm u}^{(n-1)} $,
so that the Rydberg excitation should be unblocked (which is
denoted by subscript `u').
Next, at the smaller distance~$ R_{\rm b}^{(n-1)} $ this excitation
will be blocked again, since the corresponding energy level
comes out of the irradiation bandwidth.
The same effect should take place evidently for the split sublevels
of the state $ | n\!+\!1 \rangle $, unblocking the Rydberg excitation
in the range of distances~$ R_{\rm u}^{(n+1)} $
and~$ R_{\rm b}^{(n+1)} $.

As a result, along with the commonly considered excitation zone
$ r \geqslant R_{\rm b}^{(n)} $, the additional excitation zones
[$ R_{\rm b}^{(n-1)} $, $ R_{\rm u}^{(n-1)} $],
[$ R_{\rm b}^{(n+1)} $, $ R_{\rm u}^{(n+1)} $],
[$ R_{\rm b}^{(n-2)} $, $ R_{\rm u}^{(n-2)} $],
[$ R_{\rm b}^{(n+2)} $, $ R_{\rm u}^{(n+2)} $],
\textit{etc.} will emerge (thick strips near the left-hand part of
the horizontal axis in figure).
Therefore, domain of the Rydberg blockade (corresponding to
the radii beyond the above-mentioned strips) will represent
a number of co-centric spherical shells rather than a solid ball.
It is quite surprising that possibility of emergence of
this fine structure was not taken into account before.

Let us perform some quantitative estimates.
For simplicity, we shall consider a dipolar electric field
by the central Rydberg atom averaged over orientation
of the dipole:
\begin{equation}
{\cal E}(r) =
  \frac{ C_3 e a_0 n^2 }{ r^3 } \, ,
\end{equation}
where
$ C_3 $~is a numerical coefficient on the order of unity,
$ n $~is the principal quantum number,
$ a_0 = {\hbar}^2 / ( m e^2 ) $~is Bohr radius,
$ \hbar $~is Planck constant,
$ e $ and $ m $~ are the electron charge and mass.
In the atomic units, marked by tildes, this formula can be
rewritten as
\begin{equation}
\tilde{\cal E}(\tilde{r}) =
  \frac{ C_3 n^2 }{ {\tilde{r}}^3 } \, ,
\label{eq:Electric_field_dipole}
\end{equation}
where
$ r = a_0 \tilde{r} $,
$ {\cal E} = {\cal E}^* \tilde{\cal E} $, and
$ {\cal E}^* \! = m^2 e^5 \! / {\hbar}^4 $.

Next, we shall assume that this electric field is approximately
constant at the scale of the neighboring atom, whose energy
levels are split. Such an approximation was often used in
the early works on Rydberg blockade. In fact, as follows from
the modern treatments, nonuniformity of the electric field is
essential; so that the resulting interaction between the atoms
should be of the Van der Waals type, whose energy is proportional
to~$ r^{-6} $. Unfortunately, treating the perturbations of
energy levels by the nonuniform field represents a much more
difficult task, which requires a separate paper. So, we shall
perform our estimates here ignoring the nonuniformity.

Then, under the above-mentioned assumptions, splitting the
energy levels is given by the well-known formula for
the linear Stark effect~\cite{gal94,yav80}:
\begin{equation}
\tilde{E}_{n n_1 n_2} =
  - \frac{1}{2 n^2}
  + \frac{3}{2} \, ( n_1 - n_2 ) \, n \tilde{\cal E} \, ,
\label{eq:Linear_Stark_effect}
\end{equation}
where
$ E = E^* \tilde{E} $,
$ E^* \! = m e^4 / {\hbar}^2 $,
$ n_1 $ and $ n_2 $~are the so-called parabolic quantum numbers,
which are non-negative integers, satisfying the condition:
\begin{equation}
n_1 + n_2 + |m| + 1 = n \, ,
\end{equation}
where $ m $~is the magnetic quantum number.

In particular, at $ m \! = \! 0 $ the difference of parabolic
quantum numbers can take the following values:
$ n_1 \! - \! n_2 = n \! - \! 1, \, n \! - \! 3, \: \dots \, , \,
-n \! + \! 1 $;
at $ m \! = \! 1 $,
$ n_1 \! - \! n_2 = n \! - \! 2, \, n \! - \! 4, \: \dots \, , \,
-n \! + \! 2 $;
and so on.

Let us take, for example, $ m = 0 $, which closely resembles
the real experiments on Rydberg blockade~%
\footnote{
Almost all experiments performed by now employed the energy
levels with
$ |m| \leqslant 2 \ll n $,
where $ n $~is the principal quantum number of the Rydberg state.
}.
Then, expression~(\ref{eq:Linear_Stark_effect}) can be rewritten as
\begin{equation}
\tilde{E}_{nk} \! = \frac{1}{2} \,
  \Big\{ \! - \! \frac{1}{n^2} \, + \, 3 k n \tilde{\cal E} \Big\} \, ,
\label{eq:Energy_levels}
\end{equation}
where we denoted for conciseness
\begin{equation}
k \equiv \, n_1 - n_2 \, = \,
  n \! - \! 1, \, n \! - \! 3, \, \dots , - n \! + \! 1 \, .
\label{eq:Values_k}
\end{equation}
Since Rydberg blockade is defined by the behavior of the most
disturbed energy levels, we shall consider further the sublevels
with $ k = \pm ( n \! - \! 1 ) $.

As is seen in Fig.~\ref{fig:Energy_levels}, blockade of
the basic state~$ | n \rangle $ (\textit{i.e.}, the state with
the same principal quantum number as for the central Rydberg atom)
develops under condition:
\begin{equation}
{\tilde{E}}_{n, \pm (n - 1)} \big( {\tilde{\cal E}}_{\rm b}^{(n)} \big) = \,
  {\tilde{E}}_{n, \pm (n - 1)} (0) \, \pm \,
  \frac{1}{2} \, \Delta \tilde{E} \, ,
\label{eq:Blockade_basic}
\end{equation}
where
$ {\tilde{\cal E}}_{\rm b}^{(n)} \! =
{\tilde{\cal E}} \big( {\tilde{R}}_{\rm b}^{(n)} \big) $,
and $ \Delta \tilde{E} $~is the characteristic bandwidth of
the exciting irradiation.
Substitution of~(\ref{eq:Energy_levels}) to~(\ref{eq:Blockade_basic})
results in
\begin{equation}
3 n^2 {\tilde{\cal E}}_{\rm b}^{(n)} = \, \Delta \tilde{E} \, ,
\label{eq:Condition_blockade_basic}
\end{equation}
where we have ignored the corrections of order~$ n^{-1} $.

Next, let us consider formation of the additional excitation zone
[$ R_{\rm b}^{(n-1)} $, $ R_{\rm u}^{(n-1)} $]
due to the entrance of a strongly disturbed sublevel of
the low-lying state~$ | n \! - \! 1 \rangle $ into
the band of exciting irradiation.
As follows from~(\ref{eq:Values_k}), the corresponding sublevel
should have $ k = (n-1) - 1 = n - 2 $.
Therefore, the required condition can be formulated as
\begin{equation}
{\tilde{E}}_{n-1, n-2} \big( {\tilde{\cal E}}_{\rm u, b}^{(n-1)} \big) = \,
  {\tilde{E}}_n (0) \, \mp \,
  \frac{1}{2} \, \Delta \tilde{E} \, ,
\label{eq:Blockade_additional}
\end{equation}
where minus sign refers to the situation when the perturbed sublevel
passes the lower boundary of the excitation band (\textit{i.e.},
the Rydberg excitation is unblocked), and plus sign refers to
the situation when this sublevel passes the upper boundary
(\textit{i.e.}, the Rydberg excitation is blocked again).
Substitution of~(\ref{eq:Energy_levels}) to~(\ref{eq:Blockade_additional})
results in
\begin{equation}
3 n^5 {\tilde{\cal E}}_{\rm u, b}^{(n-1)} = \,
  2 \, \mp \, n^3 \Delta \tilde{E} \, ,
\label{eq:Condition_blockade_additional}
\end{equation}
where terms of the order~$ n^{-1} $ were ignored again.

At last, combining expressions~(\ref{eq:Condition_blockade_basic})
and~(\ref{eq:Condition_blockade_additional}) we arrive at the following
equality:
\begin{equation}
3 n^5 \big[ {\tilde{\cal E}}_{\rm u, b}^{(n-1)} \pm \,
  {\tilde{\cal E}}_{\rm b}^{(n)} \big] = \, 2 \, ,
\label{eq:Blockade_final_relation}
\end{equation}
which gives a relation between the electric field strength
in the main and additional zones of Rydberg excitation.

Substitution of formula~(\ref{eq:Electric_field_dipole})
for the average dipolar electric field produced by the central
Rydberg atom into~(\ref{eq:Blockade_final_relation})
enables us to find the boundaries of the additional excitation zone:
\begin{equation}
\tilde{R}_{\rm u, b}^{(n-1)} = \,
  {\bigg( \! \frac{ 3 C_3 n^7 }{2} \! \bigg)}^{\!\! 1/3}
  {\Bigg\{ 1 \mp \, \frac{ 3 C_3 n^7 }{ 2 {\big( \tilde{R}_{\rm b}^{(n)}
  \big)}^{\! 3} } \Bigg\}}^{\! -1/3}
\label{eq:Boundaries_add_zone}
\end{equation}
or, if the second term is small as compared to the first one,
\begin{equation}
\tilde{R}_{\rm u, b}^{(n-1)} \approx \,
  {\bigg( \! \frac{ 3 C_3 n^7 }{2} \! \bigg)}^{\!\! 1/3}
  \Bigg\{ 1 \pm \, \frac{ C_3 n^7 }{ 2 {\big( \tilde{R}_{\rm b}^{(n)}
  \big)}^{\! 3} } \Bigg\} \, .
\label{eq:Boundaries_add_zone_approx}
\end{equation}

A similar analysis can be performed for split sublevels of
the higher-lying state~$ | n \! + \! 1 \rangle $.
The most disturbed sublevel in this case will have
$ k = -(n+1) + 1 = -n $.
Although the intermediate formulas are somewhat different,
the final result in the first approximation turns out to be
exactly the same as~(\ref{eq:Boundaries_add_zone})
and~(\ref{eq:Boundaries_add_zone_approx}).
(However, expressions for $ \tilde{R}_{\rm u, b}^{(n-1)} $
and $ \tilde{R}_{\rm u, b}^{(n+1)} $ will be slightly
different if terms of the order~$ n^{-1} $ are taken into
account. This fact is schematically shown in
Fig.~\ref{fig:Energy_levels}.)

Therefore, as follows from the above-written formulas,
center of the additional excitation zone is located at
the distance
\begin{equation}
\tilde{R}_{\rm c}^{(n-1)} =
  {\bigg( \! \frac{ 3 C_3 n^7 }{2} \! \bigg)}^{\!\! 1/3} ,
\label{eq:Center_add_zone}
\end{equation}
and its characteristic width equals
\begin{equation}
\Delta \tilde{R}^{(n-1)} =
  {\bigg( \frac{3}{2} \bigg)}^{\! \! 1/3}
  \frac{ C_3^{4/3} n^{28/3} }{{\big( \tilde{R}_{\rm b}^{(n)}
  \big)}^{\! 3}} \: .
\label{eq:Width_add_zone}
\end{equation}
As should be expected, position of the additional zone is
function of only the principal quantum number
(\textit{i.e.}, location of the energy level), while
its width depends also on the radius of the main blockade zone.
Really, decrease in $ \tilde{R}_{\rm b}^{(n)} $ implies
a broader excitation band $ \Delta \tilde{E} $ and, therefore,
a wider additional excitation zone.

Let us perform the numerical estimates for a particular
experiment, \textit{e.g.}~\cite{sch12}, where
$ n \! = 43 $ and $ R_{\rm b}^{(n)} \! = 4\,{\mu}{\rm m} $
($ \tilde{R}_{\rm b}^{(n)} \! = 7.6{\times}10^4 $).
Then, assuming that the numerical factor~$ 3 C_3 / 2 $ is
about unity, we get
$ \tilde{R}_{\rm c}^{(n-1)} \!\! \approx 6.5{\times}10^3 $ or
$ R_{\rm c}^{(n-1)} \!\! \approx 0.34\,{\mu}{\rm m} $.
In fact, the pair correlation function of Rydberg atoms
presented in Fig.~3a of paper~\cite{sch12} really has a strong
unexpected maximum at the radii~$ r \lesssim 0.5\,{\mu}{\rm m} $,
\textit{i.e.}, well inside the expected Rydberg blockade zone.
However, this maximum was attributed by the authors to
the imperfection of the detection procedures.
On the other hand, as follows from the above-written estimates,
this extra peak could have a deep physical meaning. Namely,
it might be caused just by the excitation of Rydberg states
with neighboring principal quantum numbers~$ n \! - \! 1 $,
$ n \! + \! 1 $, \textit{etc}.

As regards a characteristic width of the additional excitation
zone, formula~(\ref{eq:Width_add_zone}) gives
$ \Delta \tilde{R}^{(n-1)} \! \sim 5{\div}10 $,
\textit{i.e.}, only about a typical size of the ground-state atom.
However, it should be kept in mind that this excitation zone
is actually formed by a very large number ($ \sim n^2 $) of
the split sublevels, which sequentially enter and leave the
energy band~$ \Delta E $ in Fig.~\ref{fig:Energy_levels}.
So, the overall width of the additional excitation zone should be
substantial.

In conclusion, let us mention that the effect under consideration
may be important also in the experiments on the formation of
ultracold plasmas from Rydberg gases
(\textit{e.g.},~\cite{gou01,ber03,dum11}).
Namely, one of the well-known phenomena is a spontaneous
ionization of the neutral Rydberg gas, which
usually requires a number of seed electrons for the ionization
avalanche to develop~\cite{rob13}.
It can be conjectured that just the emergence of the additional
excitation zones at the small interparticle separation and
the resulting reconfiguration of the electric field pattern
lead to the inherent instability of the system and the release
of free electrons.

In summary, we have shown that the Rydberg blockade zone
should represent a number of co-centric shells rather than
a solid sphere, as it was commonly assumed before.
In fact, the presence of such inner shells, manifesting
themselves as an additional peak in the pair correlation
function at small distances, was already seen in some
experiments. Emergence of the additional excitation zones
can, firstly, reduce an overall efficiency of the Rydberg
blockade and, secondly, result in the release of some number
of free electrons. Such effects should be taken into account
in the design and interpretation of future experiments.

\acknowledgments

I am grateful to Prof. A.~Buchleitner
for valuable discussions, consultations, and advises.


\end{document}